# Capturing Aerodynamic Characteristics of ATTAS Aircraft with Evolving Intelligent System

Aydoğan Soylu
*Control and Automation Engineering Department*
*Istanbul Technical University*
Istanbul, Turkey
soylu17@itu.edu.tr

Tufan Kumbasar
*AI and Intelligent Systems Laboratory*
*Istanbul Technical University*
Istanbul, Turkey
kumbasart@itu.edu.tr

*Abstract*— Accurate modeling of aerodynamic coefficients is crucial for understanding and optimizing the performance of modern aircraft systems. This paper presents the novel deployment of an Evolving Type-2 Quantum Fuzzy Neural Network (eT2QFNN) for modeling the aerodynamic coefficients of the ATTAS aircraft to express the aerodynamic characteristics. eT2QFNN can represent the nonlinear aircraft model by creating multiple linear submodels with its rule-based structure through an incremental learning strategy rather than a traditional batch learning approach. Moreover, it enhances robustness to uncertainties and data noise through its quantum membership functions, as well as its automatic rule-learning and parameter-tuning capabilities. During the estimation of the aerodynamic coefficients via the flight data of the ATTAS, two different studies are conducted in the training phase: one with a large amount of data and the other with a limited amount of data. The results show that the modeling performance of the eT2QFNN is superior in comparison to baseline counterparts. Furthermore, eT2QFNN estimated the aerodynamic model with fewer rules compared to Type-1 fuzzy counterparts. In addition, by applying the Delta method to the proposed approach, the stability and control derivatives of the aircraft are analyzed. The results prove the superiority of the proposed eT2QFNN in representing aerodynamic coefficients.

## I. Introduction

Accurate mathematical models of aircraft dynamics are crucial for tasks like flight control design and understanding aircraft behavior. Aerodynamic forces and moments significantly influence these dynamics and make aerodynamic modeling essential. While wind tunnel experiments or parameter estimation methods are commonly used, wind tunnel testing is costly due to the required experimental setup. Therefore, many statistically based methodologies have been developed in the literature for estimating aerodynamic control and stability derivatives using measured flight tests [1]-[4]. Among these statistical methods, output error and filter error methods are widely used [1]-[3], [5]. Also, Ordinary Least Square (OLS) is used for estimating aerodynamic parameter derivatives [6]. In [7], the aerodynamic derivatives were estimated and validated using equation error and output error methods applied to simulated F-16 test data.

Neural Networks (NNs) have demonstrated significant success in aerodynamic modeling [8]-[10], yet, their interpretability remains limited. On the other hand, Fuzzy Inference Systems (FISs) are highly effective for system identification, offering excellent interpretability while adeptly modeling complex and nonlinear systems. Therefore, researchers have also addressed modeling the aerodynamics of ATTAS aircraft using FISs and demonstrated the high accuracy of the model through validation tests [11]−[13]. In another study, aerodynamic parameters were estimated using a new Adaptive Neuro Fuzzy Inference System (ANFIS) and showed that it can successfully derive aerodynamic derivatives [14].

In this paper, we present the design and deployment of the Evolving Type-2 Quantum Fuzzy Neural Network (eT2QFNN) to model the aerodynamic coefficients of the ATTAS aircraft with real flight data. This evolving structure, which contains linguistic and NN structures, uses the flight data in the train set to learn the rule structure defined with Quantum Membership Functions (QMFs) that can represent the nonlinear characteristics of the aircraft. Thus, the nonlinear characteristic of the aircraft is expressed by the linear model and QMF corresponding to each rule. The resulting rule structure also makes it possible to interpret the estimated aerodynamic model. Furthermore, adapted QMF addresses the inefficiency and noise issues in traditional membership functions of fuzzy sets. The uncertain jump positions of QMFs determine the uncertainties in the inputs of the aerodynamic model, resulting in a highly accurate model. We estimated the aerodynamic model for six coefficients of the ATTAS aircraft using two settings: (1) 80% training and 20% testing, and (2) 50% training and 50% testing. This enabled us to analyze models trained with both large and small datasets. Subsequently, we compared the aerodynamic models from eT1QFNN, eT2QFNN, OLS, ANFIS, and NN in a simulation environment to demonstrate the efficiency of the proposed eT2QFNN. We also show that eT2QFNN more accurately estimates the aerodynamic model with a smaller rule base than eT1QFNN and ANFIS. Moreover, we use the Delta method on eT2QFNN to derive stability and control derivatives of the aerodynamic model and show their consistency with the characteristics of the aircraft.

The paper is organized as follows: Section II covers the aerodynamic model of the ATTAS aircraft, Section III describes the eT2QFNN architecture, Section IV details the training method, Section V presents the comparative performance analysis, and Section VI concludes the paper.

## II. Aerodynamic Model of ATTAS Aircraft

To begin the system identification procedure, the actual aerodynamic coefficients must be obtained for comparison with the estimated ones. The calculation of flight-derived aerodynamic coefficients for the ATTAS aircraft is detailed in [1]. The aerodynamic model should balance parameter sufficiency with the risk of overfitting. Using the stepwise algorithm and analyzing flight dynamics, we define the following generic linear model for ATTAS [1], [3]:

$$C_D = C_{D_0} + C_{D_\alpha}\alpha + C_{D_{q_n}}q_n + C_{D_{\delta_e}}\delta_e \quad (1)$$

$$C_L = C_{L_0} + C_{L_\alpha}\alpha + C_{L_{q_n}}q_n + C_{L_{\delta_e}}\delta_e \quad (2)$$

$$C_M = C_{M_0} + C_{M_\alpha}\alpha + C_{M_{q_n}}q_n + C_{M_{\delta_e}}\delta_e \quad (3)$$

$$C_Y = C_{Y_0} + C_{Y_\beta}\beta + C_{Y_{p_n}}p_n + C_{Y_{r_n}}r_n + C_{Y_{\delta_r}}\delta_r \quad (4)$$

$$C_R = C_{R_0} + C_{R_\beta}\beta + C_{R_{p_n}}p_n + C_{R_{r_n}}r_n + C_{R_{\delta_a}}\delta_a \quad (5)$$

$$C_N = C_{N_0} + C_{N_\beta}\beta + C_{N_{p_n}}p_n + C_{N_{r_n}}r_n + C_{N_{\delta_r}}\delta_r \quad (6)$$

Here, $C_D, C_L, C_M, C_Y, C_R$ and $C_N$ denote the drag force, lift force, pitching moment, side force, rolling moment and yawing moment coefficients, respectively. $\alpha$, $\beta$, $q_n$, $p_n$, $r_n$ $\delta_e, \delta_a$ and $\delta_r$ are the angle of attack, side slip angle, normalized pitch rate, normalized roll rate, normalized yaw rate, elevator deflection, aileron deflection and rudder deflection inputs, respectively. $C_{D_0}, C_{D_\alpha}, C_{D_{q_n}}, C_{L_0}, C_{L_\alpha}, C_{L_{q_n}}$, $C_{M_0}, C_{M_\alpha}, C_{M_{q_n}}, C_{Y_0}, C_{Y_\beta}, C_{Y_{p_n}}, C_{Y_{r_n}}, C_{R_0}, C_{R_\beta}, C_{R_{p_n}}, C_{R_{r_n}}, C_{N_0}$, $C_{N_\beta}, C_{N_{p_n}}, C_{N_{r_n}}$ and $C_{D_{\delta_e}}, C_{L_{\delta_e}}, C_{M_{\delta_e}}, C_{Y_{\delta_r}}, C_{R_{\delta_a}}, C_{N_{\delta_r}}$ are stability and control derivatives, respectively.

### III. eT2QFNN Architecture

eT2QFNN architecture including quantum fuzzy sets comprises a multi-input-single-output network topology with five layers. $I$ denotes input features; $M$ is the output nodes and $K$ presents the term nodes for each input feature. The rule structure originates from the Interval Type-2 QMF(IT2QMF) [15]-[17], with fuzzy rules defined in (7) for $K$ term nodes.

$$R_j = \text{If } x_1 \text{ is } \tilde{Q}_{1j} \text{ and } x_i \text{ is } \tilde{Q}_{iK}, \text{ Then, } y_j = X_e \tilde{\Omega}_j \quad (7)$$

where $x_i$ is the $i$th input, $y_j$ is the output in the $j$th rule, $\tilde{Q}_{ij} = [\overline{Q}_{ij}, \underline{Q}_{ij}]$ is the IT2QMF that is defined with upper and lower QMFs that are parameterized with $\tilde{\Omega}_j = [\overline{\Omega}_j, \underline{\Omega}_j]$, $\tilde{\Omega}_j \in \mathbb{R}^{M \times 2(I+1)}$. The IT2QMF is expanded into an IT2MF with uncertain jump functions in order to detect input uncertainties [17]. The mathematical expression of the QMF for $i$th input and $j$th rule is given as follows:

$$\tilde{Q}_{ij}(x_{ij}, \gamma, m_{ij}, \tilde{\theta}_{ij}) =$$
$$\frac{1}{n_s}\sum_{r=1}^{n_s}\left[\left(\frac{1}{1+\exp(-\gamma x_i - m_{ij} + |\tilde{\theta}_{ij}^r|)}\right)U(x_i; -\infty, m_{ij})\right.$$
$$\left.+ \left(\frac{\exp(-\gamma(x_i - m_{ij} - |\tilde{\theta}_{ij}^r|))}{1+\exp(-\gamma(x_i - m_{ij} - |\tilde{\theta}_{ij}^r|))}\right)U(x_i; m_{ij}, \infty)\right] \quad (8)$$

where $m_{ij}$, $\gamma$, and $n_s$ represent mean of $i$th input feature in the $j$th rule, slope factor, and number of grades, respectively. $\tilde{\theta}_{ij} = [\overline{\theta}_{ij} \ \underline{\theta}_{ij}]$ is the set of uncertain jump positions, which is defined as $\tilde{\theta}_{ij} \in \mathbb{R}^{2 \times I \times n_s \times K}$. The upper and lower jump position is denoted as $\overline{\theta}_{ij} = [\overline{\theta}_{1j}^1 \dots \overline{\theta}_{1j}^{n_s}; \dots; \overline{\theta}_{Ij}^1 \dots \overline{\theta}_{Ij}^{n_s}]$ and $\underline{\theta}_{ij} = [\underline{\theta}_{1j}^1 \dots \underline{\theta}_{1j}^{n_s}; \dots; \underline{\theta}_{Ij}^1 \dots \underline{\theta}_{Ij}^{n_s}]$, which is expressed as $\overline{\theta}_{ij} > \underline{\theta}_{ij}$. The lower and upper jump positions of the QMF are shown in Fig. 1. An IT2 inference method is produced by the output of IT2QMF and yields a footprint of uncertainty (FOU). Furthermore, the union function $U$ with $U(x; a, b) = \begin{cases} 1, & \text{if } a \leq x_i < b \\ 0, & \text{otherwise} \end{cases}$ means that $\dot{x}_i, m_{ij}, \infty$ exhibit the properties of commutativity, associativity, monotonic, and continuous [18].

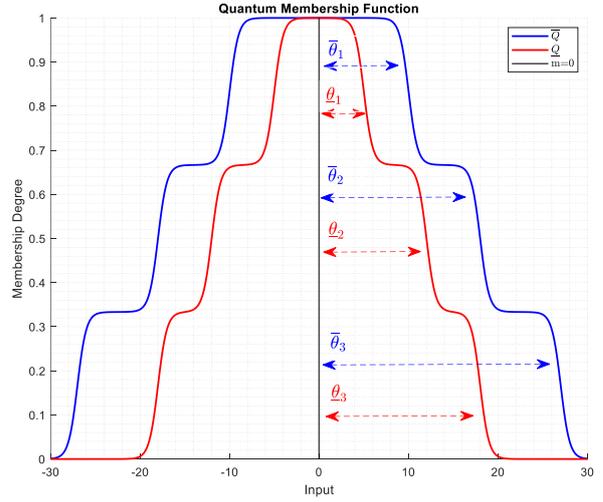

Fig. 1. QMF with upper and lower jump positions which $n_s = 3$

The eT2QFNN is defined with the following layers [17].

*Layer 1 (Input Layer):* Input signals are transmitted to this layer, passing directly to the next without computation where the $n$th observation, $x_i$ is defined as $X_n \in \mathbb{R}^{1 \times I}$.

$$u_i = x_i \quad (9)$$

*Layer 2 (Quantum Layer):* This layer performs the fuzzification procedure where QMFs determine $X_n$'s degree of membership in each rule. $K$ is the total number of defined rules and the quantum layer outputs are given in (10).

$$\overline{Q}_{ij} = \tilde{Q}_{ij}(x_i, \gamma, m_{ij}, \overline{\theta}_{ij}) \text{ and } \underline{Q}_{ij} = \tilde{Q}_{ij}(x_i, \gamma, m_{ij}, \underline{\theta}_{ij}) \quad (10)$$

*Layer 3 (Rule Layer):* The firing strength of all rules is calculated as follows:

$$\tilde{R}_j = [\overline{R}_j, \underline{R}_j] \quad (11)$$

where $\tilde{R}_j$ presents the firing strength. $\overline{R}_j = [\overline{R}_1 \dots \overline{R}_K]$ and $\underline{R}_j = [\underline{R}_1 \dots \underline{R}_K]$ are the upper firing strength and lower firing strength, respectively. Also, the product T-norm is used with the IT2QMF, as defined in (12).

$$\overline{R}_j = \mathbf{\Pi}_{i=1}^I \overline{Q}_{ij} \text{ and } \underline{R}_j = \mathbf{\Pi}_{i=1}^I \underline{Q}_{ij} \quad (12)$$

*Layer 4 (Output Processing Layer):* The following type reduction is performed using the design factor $q$ to convert type 2 variables to type 1 variables [19]:

$$y_u = (1-q)\frac{\sum_{j=1}^K \overline{R}_j \ \overline{\Omega}_j x_e^T}{\sum_{j=1}^K \overline{R}_j}, \ y_l = q\frac{\sum_{j=1}^K \underline{R}_j \ \underline{\Omega}_j x_e^T}{\sum_{j=1}^K \underline{R}_j} \quad (13)$$

where $\overline{\Omega}_j$ and $\underline{\Omega}_j$ denote upper and lower the layer weight parameter for the $j$th rule and $\mathbf{x}_e = [1, x_1, \dots, x_I]$ is the extended input vector.

*Layer 5 (Output Layer):* The crisp output is as follows:

$$y_{out} = y_u + y_l \quad (14)$$

## IV. Training Method of eT2QFNN

The rule growth criteria and parameter update scenarios are the two scenarios that train the eT2QFNN, which operates on an online learning policy [17].

### A. Rule Growing Mechanism

The learning process starts with an empty rule base and the structure is continuously updated as a new data sample arrives. With each new data sample, eT2QFNN generates a hypothetical rule that drives the autonomous evolution of its fuzzy rules. The generated hypothetical rules need to evolve significantly before being incorporated into the network structure. To predict complex changes in the data, the significance is evaluated using a Gaussian Mixture Model (GMM) [20] which can be expressed as

$$\hat{E}_{sig}(j) = \|\omega_j\| \left( \pi^{\frac{I}{2}} \det(\Sigma_j)^{\frac{1}{2}} N_j A^T \right)^{1/2} \quad (15)$$

where $\Sigma_j = \text{diag}(\sigma_{1,j}^2, \dots, \sigma_{I,j}^2)$ denotes the positive definite weighting matrix, $A = [\alpha_1, \dots, \alpha_H]$ presents GMM mixing coefficients vector with Gaussian coefficients $\alpha_1, \dots, \alpha_H$ and $\omega_j$ is the vector of rule $j$ and $N_J$ is defined as follows:

$$N_J = \left[ N\left(m_j - v_1; 0, \frac{\Sigma_j}{2} + \Sigma_1\right), N\left(m_j - v_2; 0, \frac{\Sigma_j}{2} + \Sigma_2\right), \dots, N\left(m_j - v_H; 0, \frac{\Sigma_j}{2} + \Sigma_H\right) \right] \quad (16)$$

Here, $m_j = [m_{1,j}, \dots, m_{I,j}]$ presents the mean vector of the $j$th rule. $v_H \in \mathbb{R}^I$ is the variance matrix and $H$ denotes the number of components.

eT2QFNN uses IT2QMF rather than the Gaussian membership function so (15) is not immediately applicable to eT2QFNN rule significance estimation. The main solution to this issue is to use the IT2 Gaussian MF (IT2GMF) to approximate IT2QMF [17]. The mathematical expression of this approach can be expressed as follows:

$$\tilde{Q}_{i,j}(x_i, \gamma, m_{i,j}, \tilde{\theta}_{i,j}) \approx \tilde{\mu}_{i,j} = \exp\left(-\frac{(x_i - m_{i,j})^2}{\tilde{\sigma}_{i,j}}\right) \quad (17)$$

where $\tilde{\sigma}_{i,j} = [\underline{\sigma}_{i,j}, \overline{\sigma}_{i,j}]$ with $\underline{\sigma}_{i,j} = \min \underline{\theta}_{i,j}$ and $\overline{\sigma}_{i,j} = \min \overline{\theta}_{i,j}$. Both the means of IT2GMF and IT2QMF are defined as equal and the minimum value of $\tilde{\theta}_{i,j}$ is used to define the width of upper and lower IT2GMF. An accurate approximation of IT2QMF can be obtained because the IT2GMF falls within the boundaries of IT2QMF [15], [17].

The rule significance can be estimated using the modified IT2QMF estimation, which can be expressed in (18) and the calculation of rule significance is described in [17].

$$\hat{E}_j = |\hat{E}_{j,u}| + |\hat{E}_{j,l}|. \quad (18)$$

When the hypothetical rule contributes more than the current rules, it is added to the network as a new rule $R_{K+1}$. This condition for rule growth is expressed as $E_{K+1} \geq \rho \sum_{j=1}^{K} E_j$ with the vigilance parameter $\rho \in (0,1]$ [17].

### B. Parameter Update Mechanism

A hypothetical rule is added to the network when the conditions are met, called fuzzy rule initialization. Otherwise, the network parameters are adjusted based on input changes, known as the winning rule update strategy.

*1) Initializing Fuzzy Rules:* The growing criterion calculates the importance of the hypothetical rule based on changes in the input data. At the $n$th time step, $X_n$ is assigned a new mean to IT2QMF: $m_{K+1} = X_n$. This results in a new jump position, enabled by the distance-based formulation. Also, the distance is updated by applying the mixed mean of the GMM $\hat{v}$, and the updated jump positions are given in (19).

$$\overline{\theta}_{i,1} = \frac{1}{\left(\frac{n_s + 1}{2}\right)} \cdot r \cdot \overline{\sigma}_{i,1} \text{ and } \underline{\theta}_{i,1} = \frac{1}{\left(\frac{n_s + 1}{2}\right)} \cdot r \cdot \underline{\sigma}_{i,1} \quad (19)$$

The eT2QFNN uses a diagonal positive definite matrix of mixed variance since the GMM approximates the mean and variance for complex input changes, defined as follows:

$$\overline{\sigma}_{i,1} = \hat{\sigma}_i, \quad \underline{\sigma}_{i,1} = \delta_1 \overline{\sigma}_{i,1}$$
$$\hat{\Sigma} = \sum_{h=1}^{H} \Sigma_h \cdot v_h, \quad \hat{\Sigma} = \text{diag}(\hat{\sigma}_1^2, \dots, \hat{\sigma}_I^2) \quad (20)$$

where $\delta_1$ creates the FOU. The hypothetical rule's consequent rule parameters resemble the winning rule provided by (21). Finally, when the network structure's requirement is satisfied by the hypothetical rule, a new rule $R_{K+1}$ is added, and its covariance matrix is calculated in (22). Moreover, the covariance matrix must be modified to accept new rules [21] by multiplying $(K^2 + 1)/K^2$ with the covariance matrices of the other rules, as defined in (23).

$$\tilde{\Omega}_{K+1} = \tilde{\Omega}_{jw} \quad (21)$$

$$P_{K+1}(n) = I_{Z \times Z} \quad (22)$$

$$P_j(n) = \left(\frac{K^2 + 1}{K^2}\right) P_j(n-1). \quad (23)$$

*2) Winning Rule Update:* In case the hypothetical rules fail to meet the requirements of the network, the performance of eT2QFNN must be maintained by changing the parameters of the network in response to changes in the input data. To do this, a winning rule is developed that depends on the maximum spatial firing power. The spatial firing strength evaluates the satisfaction rate of the rule's antecedent part based on input data changes [17] and is obtained as follows:

$$j_\omega = \arg \max_{j=1,\dots,K} \hat{P}(R_j \backslash X) \quad (24)$$

$$\tilde{R}_j = \left(\overline{R_j} + \underline{R_j}\right)/2. \quad (25)$$

In addition, $\overline{\Omega}, \underline{\Omega}, q, m, \overline{\theta}$ and $\underline{\theta}$ are updated via the decoupled extended Kalman filter and $q$ is passed through the sigmoid function so that $q \in (0,1)$ after updating the value of $q$. Thus, the upper and lower crisp outputs continuously adapt to changes in the inputs thanks to the decoupled extended Kalman filter [22]. This aids in the clustering of the local parameters, which forms the block diagonal covariance matrix $P(n)$ [21]. The single block covariance matrix $P_{j_\omega}(n)$ is updated for each time step. The remaining mathematical expression for the decoupled extended Kalman filter can be calculated in (26)−(29).

$$\tilde{P}(n) = \text{diag}(P_1(n), P_2(n), \ldots, P_j(n), \ldots, P_K(n)) \quad (26)$$

$$G_{j_\omega}(n) = P_{j_\omega}(n-1)H_{j_\omega}(n) \times [\eta I_{M \times M} + H_{k_\omega}^T(n)P_{j_\omega}(n-1)H_{j_\omega}(n)]^{-1} \quad (27)$$

$$P_{j_\omega}(n) = [I_{Z \times Z} - G_{j_\omega}(n)H_{j_\omega}^T(n)]P_{j_\omega}(n-1) \quad (28)$$

$$\vec{\varphi}_{j_\omega}(n) = \vec{\varphi}_{j_\omega}(n-1) + G_{j_\omega}(n)[t(n) - y(n)] \quad (29)$$

where $G_{j_\omega}(n)$ is the Kalman gain matrix, $H_{k_\omega}$ is the Jacobian matrix, $t(n)$ is the target vector and $y(n)$ is output vector at $n$th iteration and $\vec{\varphi}_{j_\omega}(n)$ is the parameter vector defined as:

$$\vec{\varphi}_{j_\omega}(n) = [\underline{\Omega}_{j_\omega}^T \quad \bar{\Omega}_{j_\omega}^T \quad q_{j_\omega}^T \quad m_{j_w}^T \quad \underline{\theta}_{j_\omega}^T \quad \bar{\theta}_{j_\omega}^T]^T \quad (30)$$

eT2QFNN, for $\bar{\theta}_{ij} = \underline{\theta}_{ij}$ and $q = 0.5$, reduces to its type-1 counterpart, the Evolving Type-1 Quantum Fuzzy Neural Network (eT1QFNN). Its parameter vector is as follows:

$$\vec{\emptyset}_{j_\omega}(n) = [\Omega_{j_\omega}^T \quad m_{j_\omega}^T \quad \theta_{j_\omega}^T]^T \quad (31)$$

## V. COMPARATIVE PERFORMANCE ANALYSIS

Here, a system identification study has been performed in a Matlab environment for (1)–(6) aerodynamic model using the pre-recorded and pre-processed short period, bank to bank, and dutch roll maneuvers of the ATTAS aircraft [23]. Both longitudinal and lateral and directional modes of the aircraft are triggered with these maneuvers and 6 coefficients of arodynamic models are estimated via eT1QFNN, eT2QFNN, OLS, ANFIS, and NN. The learning rates of eT1QFNN, eT2QFNN, NN and ANFIS are selected appropriately based on the flight data while 2 Gauss membership functions are used for each input in ANFIS. The selection of tuning parameters of NN is explained in detail in [1]. In eT1QFNN and eT2QFNN, the rule growth parameter is set as $\rho = 0.65$ while the FOU parameter of the eT2QFNN is set as $\delta_1 = 0.8$.

Theil's Inequality Coefficient (TIC) performance criterion is used to evaluate the performance of the estimation algorithms because the TIC coefficient is an indicator that evaluates the quality of the estimated model using real system data. The outputs of the calculated model are consistent with the outputs from the actual flight when the TIC value is less than 0.35. The TIC criterion is calculated as follows [24]:

$$TIC = \frac{\sqrt{\sum_{i=1}^{N}(z_{out}(i) - y_{out}(i))^2}}{\sqrt{\sum_{i=1}^{N}z_{out}(i)^2} + \sqrt{\sum_{i=1}^{N}y_{out}(i)^2}} \quad (32)$$

where $N$ is the total number of samples, $z_{out}$ and $y_{out}$ denote the measured response and the model output, respectively.

To train the models, two different settings were defined using the flight data, which consists of 2128 data points, to estimate the aerodynamic model.

- **Setting-1:** training is performed with the first 1702 points (80%) of the flight data and testing was performed with the last 426 points (20%).

- **Setting-2:** training is performed with the first 1064 points (50%) of these data points and testing with the remaining 1064 points (50%).

Thus, the performances of the aerodynamic models trained with a lot of data and with little data are observed by examining the TIC criterion and it is investigated whether the aerodynamic model can be obtained accurately with little data.

For each training set (Setting-1 and Setting-2), each aerodynamic model is trained, and the TIC values of each aerodynamic coefficient are presented in Tables I – II. Then, the performance of the model according to the TIC criterion is ranked for each aerodynamic coefficient. The estimation algorithm with the lowest TIC value was ranked first, while the one with the highest TIC value was ranked last. To better interpret the performance of the estimation algorithms, the rank number of these algorithms is averaged and the mean ranking of the estimation algorithm is obtained which is illustrated in Table III.

TABLE I. TESTING TIC VALUES FOR SETTING -1

| Coeffs. | OLS | eT1QFNN | eT2QFNN | ANFIS | NN |
|---|---|---|---|---|---|
| $C_L$ | 0.00341 | 0.003412 | 0.003408 | 0.00355 | 0.003594 |
| $C_D$ | 0.0162 | 0.01177 | 0.01155 | 0.01858 | 0.01189 |
| $C_M$ | 0.25887 | 0.25857 | 0.16775 | 0.2032 | 0.19474 |
| $C_Y$ | 0.05795 | 0.05817 | 0.05668 | 0.06158 | 0.08962 |
| $C_R$ | 0.23139 | 0.17433 | 0.15043 | 0.16016 | 0.29472 |
| $C_N$ | 0.11202 | 0.1172 | 0.10968 | 0.11966 | 0.11886 |

TABLE II. TESTING TIC VALUES FOR SETTING - 2

| Coeffs. | OLS | eT1QFNN | eT2QFNN | ANFIS | NN |
|---|---|---|---|---|---|
| $C_L$ | 0.00501 | 0.004975 | 0.003906 | 0.003909 | 0.004707 |
| $C_D$ | 0.01732 | 0.01341 | 0.01527 | 0.02265 | 0.01742 |
| $C_M$ | 0.32644 | 0.32608 | 0.28153 | 0.25846 | 0.22514 |
| $C_Y$ | 0.05449 | 0.0409 | 0.0455 | 0.093 | 0.05379 |
| $C_R$ | 0.26752 | 0.28506 | 0.23861 | 0.39095 | 0.23282 |
| $C_N$ | 0.149 | 0.08157 | 0.11166 | 0.1698 | 0.31176 |

According to Table I, eT2QFNN is the estimation algorithm with the lowest TIC value for each aerodynamic coefficient. Therefore, the eT2QFNN algorithm is ranked first for each aerodynamic coefficient so the average ranking of eT2QFNN is one. Also, Table II shows that the TIC value of eT2QFNN is quite low compared to other estimation algorithms, so its mean ranking is 2, as shown in Table III. According to Table III, the average ranking value of eT1QFNN is considerably lower than the average ranking value of OLS, ANFIS, and NN for both approaches since eT1QFNN is usually in the top 3 in the TIC criterion for both longitudinal and lateral directional aerodynamic coefficients. In addition, when estimating aerodynamic models with fewer data sets, the TIC values of OLS are generally higher than eT1QFNN, eT2QFNN, and NN. Therefore, the average ranking value of OLS is higher than these algorithms. When two different approaches are compared, there is no change in the average ranking value of ANFIS, while the average ranking of NN is lower in the second approach.

TABLE III. MEAN RANKING COMPARISON

| Setting-1 | | Setting-2 | |
|---|---|---|---|
| Model | Rank | Model | Rank |
| eT2QFNN | 1 | eT2QFNN | 2 |
| eT1QFNN | 3 | eT1QFNN | 2.5 |
| OLS | 3.17 | NN | 2.83 |
| ANFIS | 3.83 | OLS | 3.83 |
| NN | 4 | ANFIS | 3.83 |

As fuzzy models are rule-based structures, we also investigated their resulting number of rules. The average of the number of rules obtained with these rule-based estimation algorithms which is given in Table IV is calculated from the rules obtained from each aerodynamic coefficient estimation. Table IV indicates that eT2QFNN learned with less number of rules than other rule-based algorithms. ANFIS has the highest average number of rules. This has shown us that evolving structured algorithms can learn with fewer rules.

TABLE IV. MEAN NUMBER OF RULES OF FUZzy MODELS

| Model | Mean Number of Rules |
|---|---|
| eT2QFNN | 6.33 |
| eT1QFNN | 9.83 |
| ANFIS | 12 |

Delta method [25] is applied to obtain the parameters in the aerodynamic models estimated with eT1QFNN and eT2QFNN. Thus, each input variables $\alpha, \beta, q_n, p_n, r_n, \delta_e, \delta_a$ and $\delta_r$ are perturbed in both directions by around 1% to obtain the parameters of the longitudinal, lateral and directional aerodynamic coefficients. Fig. 2, Fig. 3, Fig. 4 and Fig. 5 show the values of the $C_{Y_\beta}, C_{L_\alpha}, C_{R_{\delta_a}}$ and $C_{N_{\delta_r}}$ for eT1QFNN and eT2QFNN throughout the flight data. In addition, we can investigate how many of these parameters take the relevant value throughout the entire flight period using histogram plots. The parameter derivatives of eT1QFNN and eT2QFNN for all aerodynamic coefficients can be calculated using this methodology and compared with OLS results, as shown in Table V.

TABLE V. ESTIMATED PARAMETERS OF THE SIX AERODYNAMIC COEFFICIENTS

| Parameter | OLS | eT1QFNN | eT2QFNN |
|---|---|---|---|
| $C_{L_\alpha}$ | 5.3137 | 5.3148 | 5.0012 |
| $C_{L_{q_n}}$ | 1.5413 | 1.6201 | 1.5811 |
| $C_{L_{\delta_e}}$ | 0.2878 | 0.2936 | 0.3702 |
| $C_{D_\alpha}$ | 0.2671 | 0.1457 | 0.2675 |
| $C_{D_{q_n}}$ | 1.9731 | 2.6584 | 1.3116 |
| $C_{D_{\delta_e}}$ | 0.1129 | 0.1391 | 0.1980 |
| $C_{M_\alpha}$ | −0.8832 | −0.8835 | −0.1024 |
| $C_{M_{q_n}}$ | −7.1838 | −7.1631 | −7.7521 |
| $C_{M_{\delta_e}}$ | −1.0895 | −1.0890 | −0.0770 |
| $C_{Y_\beta}$ | −1.0470 | −1.0457 | −1.0458 |
| $C_{Y_{p_n}}$ | 0.1723 | 0.1830 | 0.0453 |
| $C_{Y_{r_n}}$ | 0.6104 | 0.6305 | 0.6085 |
| $C_{Y_{\delta_r}}$ | 0.1909 | 0.1899 | 0.1512 |
| $C_{R_\beta}$ | −0.1027 | −0.0429 | −0.0884 |
| $C_{R_{p_n}}$ | −0.7606 | −0.3073 | −0.6330 |
| $C_{R_{r_n}}$ | 0.2336 | 0.3468 | 0.2737 |
| $C_{R_{\delta_a}}$ | −0.1925 | −0.1103 | −0.1851 |
| $C_{N_\beta}$ | 0.2539 | 0.2787 | 0.2589 |
| $C_{N_{p_n}}$ | −0.0444 | −0.0999 | −0.0142 |
| $C_{N_{r_n}}$ | −0.1357 | −0.0865 | −0.0858 |
| $C_{N_{\delta_r}}$ | −0.1438 | −0.1354 | −0.1376 |

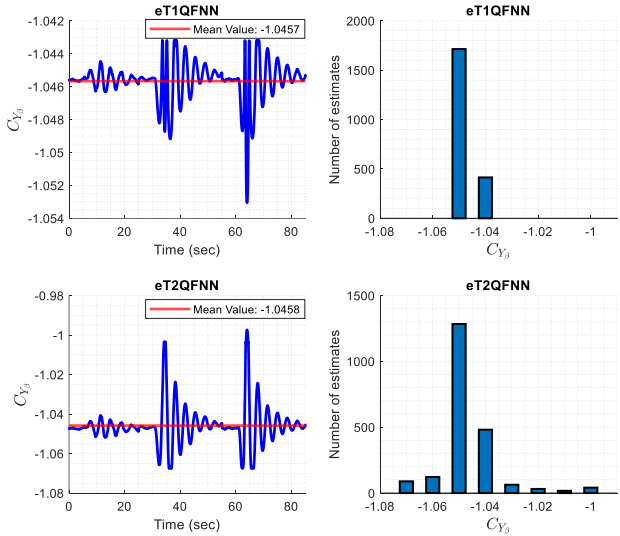

Fig. 2. $C_{Y_\beta}$ estimation of eT1QFNN and eT2QFNN using Delta Method

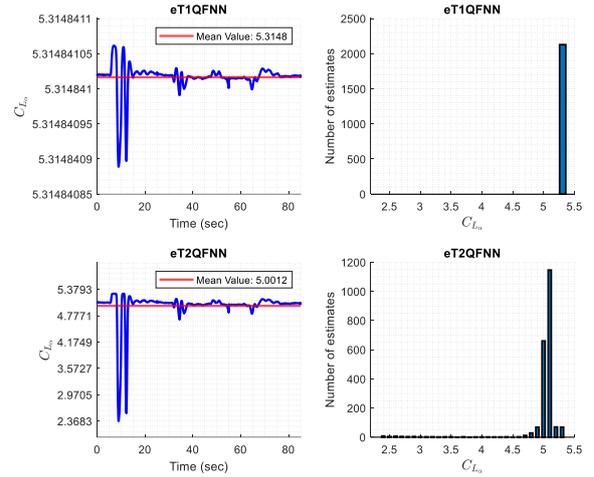

Fig. 3. $C_{L_\alpha}$ estimation of eT1QFNN and eT2QFNN using Delta Method

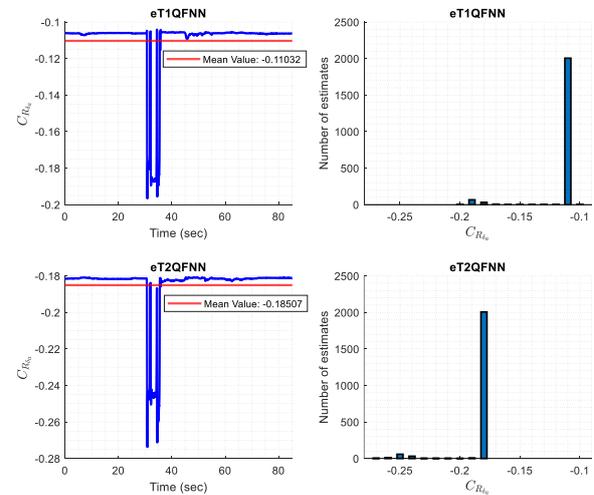

Fig. 4. $C_{R_{\delta_a}}$ estimation of eT1QFNN and eT2QFNN using Delta Method

Table V obtained from the first setting shows that the parameter derivatives estimated by eT1QFNN and eT2QFNN are consistent with OLS, accurately reflecting the stability and control behavior of the ATTAS aircraft. While OLS is simpler and sufficient for extracting linear derivatives from input-output data, eT1QFNN and eT2QFNN excel in capturing complex systems globally. Thus, we can produce models with a high degree of nonlinearity and physical understanding using evolving fuzzy systems. Furthermore, their rule base and delta method enable dynamic interpretation without solving the equations of motion.

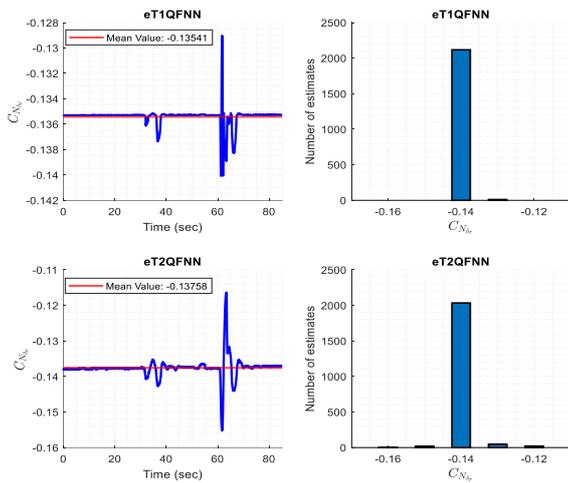

Fig. 5. $C_{N_{\delta_r}}$ estimation of eT1QFNN and eT2QFNN using Delta Method

The histogram plots in Fig. 2, Fig. 3, Fig. 4 and Fig. 5 demonstrate that eT1QFNN has less scattered parameter derivatives than eT2QFNN. Moreover, the distribution of these parameter derivatives is related to the pattern of rules learned by the evolving structures. Therefore, the parameter derivatives resulting from the perturbation of inputs in the rule structures of eT2QFNN are more sensitive.

## VI. Conclusions and Future Work

This paper presents the design and deployment of eT2QFNN to estimate the aerodynamic model of the ATTAS aircraft with real flight data. Each fuzzy rule generated by eT2QFNN calculates a localized linear model for the given input variables, which is used to model the aerodynamic control and stability derivatives. The obtained outputs of the aerodynamic model are compared with the outputs of eT1QFNN, OLS, NN and ANFIS algorithms. According to the obtained results, eT2QFNN is more successful than other algorithms in aerodynamic modeling, even with less training data. The reason for this is that eT2QFNN has QMF, automatic rule growth and parameter tuning with decoupled extended Kalman filter. In addition, these features make it more robust to noises and uncertainties in the data. Also, the delta method is integrated into the eT2QFNN to obtain the parameter derivatives in the aerodynamic model. It is concluded that these derivatives accurately represent the ATTAS aircraft's stability and characteristics.

As a future work, real-time aerodynamic modeling of a model airplane with more flight data will be performed using eT2QFNN. An adaptive control design will be developed using the obtained model.

## Acknowledgment

The authors acknowledge using ChatGPT to refine the grammar and enhance the English language expressions.